%% file: main.tex
\documentclass[journal]{elsarticle}
\usepackage[utf8]{inputenc}
\usepackage{graphicx}
\usepackage{float}
\usepackage{dblfloatfix} 
\usepackage{textcomp}
\usepackage{booktabs} 
\usepackage{multirow}
\usepackage{hyperref}
 
\graphicspath{ {./img/} }

\author[]{Matteo Pennisi\corref{corb}\fnref{fn1}}
\author[]{Isaak Kavasidis\corref{corb}\fnref{fn1}}
\author[]{Concetto Spampinato\fnref{fn1}}
\author[]{Vincenzo Schininà\fnref{fn2}}
\author[]{Simone Palazzo\fnref{fn1}}
\author[]{Francesco Rundo\fnref{fn3}}
\author[]{Massimo Cristofaro\fnref{fn2}}
\author[]{Paolo Campioni\fnref{fn2}}
\author[]{Elisa Pianura\fnref{fn2}}
\author[]{Federica Di Stefano\fnref{fn2}}
\author[]{Ada Petrone\fnref{fn2}}
\author[]{Fabrizio Albarello\fnref{fn2}}
\author[]{Giuseppe Ippolito\fnref{fn2}}
\author[]{Salvatore Cuzzocrea\fnref{fn4} }
\author[]{Sabrina Conoci\fnref{fn4}}

\cortext[corb]{Equal contribution}
\cortext[cora]{Corresponding author}

\fntext[fn1]{DIEEI, University of Catania, Catania, Italy}
\fntext[fn2]{National Institute for infectious disease, “Lazzaro Spallanzani” Department, Rome, Italy}
\fntext[fn3]{STMicroelectronics - ADG Central R\&D, Catania, Italy}
\fntext[fn4]{ChimBioFaram Department, University of Messina, Messina, Italy}
\date{November 2020}

\begin{document}
\title{An Explainable AI System for Automated COVID-19 Assessment and Lesion Categorization from CT-scans}

\begin{abstract}
\input{abstract.tex}
\end{abstract}
\maketitle

\section{Introduction}
\input{intro}

\section{Related Work}
\input{related} 

\section{Explainable AI for COVID-19 data understanding }
\input{explainable}

\section{Results and Discussion}
\input{results}

\section{Conclusions} 
\input{conclusions}

\section*{Acknowledgment}
\input{ack}

\section*{Regulation and Informed Consent}
\input{consent.tex}

\section*{Declarations of interest}
None.

\end{document}

%% file: abstract.tex
\textit{COVID-19 infection caused by SARS-CoV-2 pathogen is a catastrophic pandemic outbreak all over the world with exponential increasing of confirmed cases and, unfortunately, deaths. In this work we propose an AI-powered pipeline, based on the deep-learning paradigm, for automated COVID-19 detection and lesion categorization from CT scans. 
We first propose a new segmentation module aimed at identifying automatically lung parenchyma and lobes. Next, we combined such segmentation network with classification networks for COVID-19 identification and lesion categorization.
We compare the obtained classification results with those obtained by three expert radiologists on a dataset consisting of 162 CT scans. 
Results showed a sensitivity of 90\% and a specificity of 93.5\% for COVID-19 detection, outperforming those yielded by the expert radiologists, and an average lesion categorization accuracy of over 84\%. Results also show that a significant role is played by prior lung and lobe segmentation that allowed us to enhance performance by over 20 percent points. 
The interpretation of the trained AI models, moreover, reveals that the most significant areas for supporting the decision on COVID-19 identification are consistent with the lesions clinically associated to the virus, i.e., crazy paving, consolidation and ground glass. This means that the artificial models are able to discriminate a positive patient from a negative one (both controls and patients with interstitial pneumonia tested negative to COVID) by evaluating the presence of those lesions into CT scans.
Finally, the AI models are  integrated into a user-friendly GUI to support AI explainability for radiologists, which is publicly available at \url{http://perceivelab.com/covid-ai}. The whole AI system is unique since, to the best of our knowledge, it is the first AI-based software, publicly available, that attempts to explain to radiologists what information is used by AI methods for making decision and that involves proactively them in the decision loop to further improve the COVID-19 understanding.
}

%% file: intro.tex
At the end of 2019 in Wuhan (China) several cases of an atypical pneumonia, particularly resistant to the traditional pharmacological treatments, were observed. In early 2020, the COVID-19 virus \cite{zhu2020novel} has been identified as the responsible pathogen for the unusual pneumonia. From that time, COVID-19 has spread all around the world hitting, to date about 32 million of people (with about 1M deaths), stressing significantly healthcare systems in several countries. Since the beginning, it has been noted that 20\% of infected subjects appear to progress to severe disease, including pneumonia and respiratory failure and in around 2\% of cases death \cite{world2020novel}. 

Currently, the standard diagnosis of COVID-19 is de facto based on a biomolecular test through Real-Time Polimerase Chain Reaction (RT-PCR) test \cite{huang2020use,ng2020imaging}. However, although widely used, this biomolecular method is time-consuming and appears to be not quite accurate suffering from a large number of false-negatives \cite{liu2020clinical}. 

Recent studies have outlined the effectiveness of radiology imaging through chest X-ray and mainly Computed Tomography (CT) given the pulmonary involvement in subjects affected by the infection \cite{liu2020clinical,chung2020ct}.
Given the extension of the infection and the number of cases that daily emerge worldwide and that call for fast, robust and medically sustainable diagnosis, CT scan appears to be suitable for a robust-scale screening, given the higher resolution w.r.t. X-Ray.
In this scenario, artificial intelligence may play a fundamental role to make the whole diagnosis process automatic, reducing, at the same time, the efforts required by radiologists for visual inspection \cite{rundo2019advanced}.

In this paper, thus, we present an innovative artificial intelligent approach to achieve both COVID-19 identification and lesion categorization (ground glass, crazy and paving consolidation) that are instrumental to evaluate lung damages and the prognosis assessment. Our method relies only on radiological image data avoiding the use of additional clinical data in order to create AI models that are useful for large-scale and fast screening with all the subsequent benefits for a favorable outcome. More specifically, we propose an innovative automated pipeline consisting of 1) lung/lobe segmentation, 2) COVID-19 identification and interpretation and 3) lesion categorization. We tested the AI-empowered software pipeline on multiple CT scans, both publicly released and collected at the Spallanzani Institute in Italy, and showed that: 1) our segmentation networks is able to effectively extract lung parenchyma and lobes from CT scans, outperforming state of the art models; 2) the COVID-19 identification module yields better accuracy (as well as specificity and sensitivity) than expert radiologists. Furthermore, when attempting to interpret the decisions made by the proposed AI model, we found that it learned automatically, and without any supervision, the CT scan features corresponding to the three most common lesions spotted in the COVID-19 pneumonia, i.e., consolidation, ground glass and crazy paving, demonstrating its reliability in supporting the diagnosis by using only radiological images. 
As an additional contribution, we integrate the tested AI models into an user-friendly GUI to support further AI explainability for radiologists, which is publicly available at \url{http://perceivelab.com/covid-ai}. The GUI processes entire CT scans and reports if the patient is likely to be affected by COVID-19, showing, at the same time, the scan slices that supported the decision.

%% file: related.tex
The COVID-19 epidemic caught the scientific community flat-footed and in response a high volume of research has been dedicated at all possible levels. In particular, since the beginning of the epidemic, AI models have been employed for disease spread monitoring~\cite{allam2020coronavirus,lin2020combat,zheng2020predicting}, for disease progression \cite{bai2020predicting} and prognosis \cite{liang2020early}, for predicting mental health ailments inflicted upon healthcare workers~\cite{cosic2020artificial} and for drug repurposing~\cite{mohanty2020application,ke2020artificial} and discovery~\cite{richardson2020baricitinib}. 

However, the lion's share in employing AI models for the fight against COVID-19 belongs to the processing of X-rays and CT scans with the purpose of detecting the presence of COVID-19 or not. In fact, recent scientific literature has demonstrated the high discriminative and predictive capability of deep learning methods in the analysis of COVID-19 related radiological images\cite{brunese2020explainable,huang2020serial}. The key radiological techniques for COVID-19 induced pneumonia diagnosis and progression estimation are based on the analysis of CT and X-ray images of the chest, on which deep learning methodologies have been widely used with good results for segmentation, predictive analysis, and discrimination of patterns \cite{nardelli2018pulmonary,navab2015medical,mei2020artificial}. If, on one hand, X-Ray represents a cheaper and most effective solution for large scale screening of COVID-19 disease, on the other hand, its low resolution has led AI models to show lower accuracy compared to those obtained with CT data. 

For the above reasons, CT scan has become the gold standard for investigation on lung diseases. In particular, deep learning, mainly in the form of Deep Convolutional Neural Networks (DCNN), has been largely applied to lung disease analysis from CT scans images, for evaluating progression in response to specific treatment (for instance immunotherapy, chemotherapy, radiotherapy) \cite{setio2016pulmonary,cha2017bladder}, but also for interstitial lung pattern analysis \cite{bermejo2020classification,gao2018holistic} and on segmentation and discrimination of lung pleural tissues and lymph-nodes \cite{moltz2009advanced,shin2016deep}. This latter aspect is particularly relevant for COVID-19 features and makes artificial intelligence an extremely powerful tool for supporting early diagnosis of COVID-19 and disease progression quantification. As a consequence, several recent works have reported using AI models for automated categorization of CT scans \cite{mei2020artificial} and also on COVID-19~\cite{li2020artificial,shi2020review,bai2020ai} but without being able to distinguish between the various types of COVID-19 lesions. 

Thus, the main contributions of this paper w.r.t. the state of the art are the following ones:
\begin{itemize}
    \item We propose a novel lung-lobe segmentation network outperforming state of the art models;
    \item We employ the segmentation network to drive a classification network in first identifying CT scans of COVID-19 patients, and, afterwards, in automatically categorizing specific lesions;
    \item We then provide interpretation of the decisions made by the employed models and discover that, indeed, those models focus on specific COVID-19 lesions for distinguishing whether a CT scan pertains COVID-19 patients or not;
    \item We finally integrate the whole AI pipeline into a web platform to ease use for radiologists, supporting them in their investigation on COVID-19 disease.
\end{itemize}

%% file: explainable.tex
The proposed AI system aims at 1) extracting lung and lobes from chest CT data, 2) categorizing CT scans as either COVID-19 positive or COVID-19 negative; 3) identifying and localizing typical COVID-19 lung lesions (consolidation, crazy paving and ground glass); and 4) explaining eventually what CT slices it based its own decisions. 

\subsection{AI Model for Lung Segmentation}
\input{segm.tex}\label{ssec:segmentation}
\subsection{Automated COVID-19 Diagnosis: CT classification}

\begin{figure*}[hbt!]
    \centering
    \includegraphics[width=\textwidth]{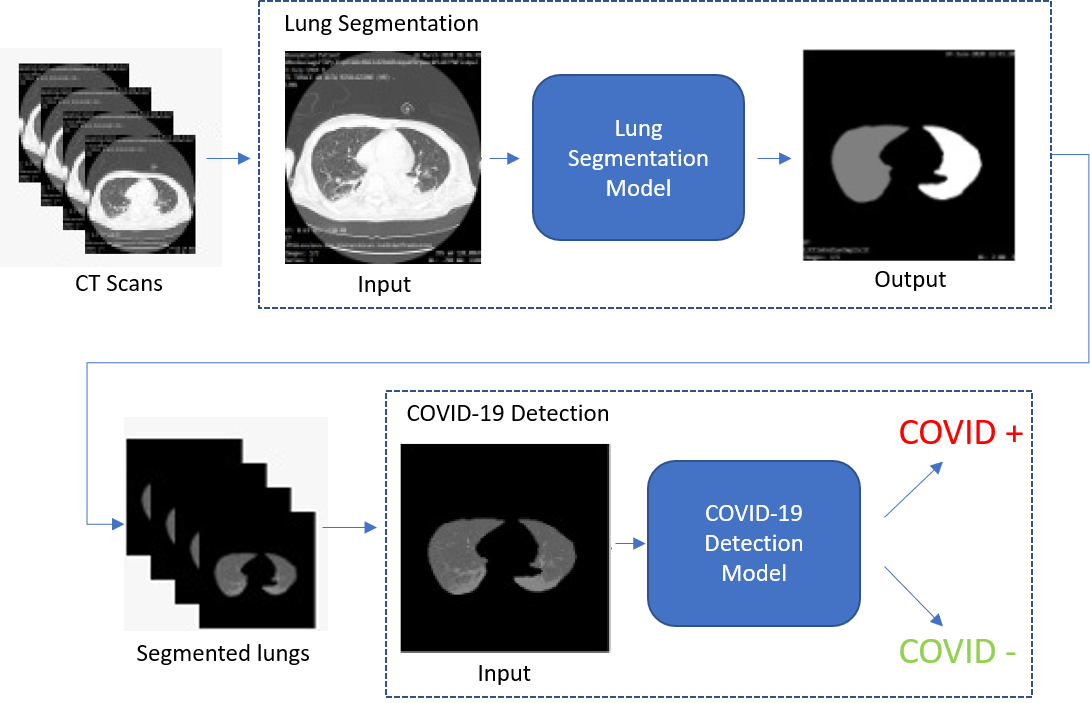}
        \centering
    \caption{Overview of the \textbf{COVID-19 detection approach} for CT scan classification as either COVID-19 positive or negative.}
    \label{fig:overview}
\end{figure*}

After parenchima lung segmentation (through the segmentation model presented in \ref{ssec:segmentation}) a deep classification model analyzes slice by slice, each segmented CT scan, and decides whether a single slice contains some evidence of the COVID-19 disease. Afterwards, a voting method provides its final prediction according to all the per-slice decisions. At this stage, the system does not carry out any identification and localization of COVID-19 lesions, but it just identifies all slices where patterns of interest may be found and according to them, makes a guess on the presence or not of COVID-19 induced infection. An overview of this model is shown in Fig.~\ref{fig:overview}: first the segmentation network, described in the previous section, identifies lung areas from CT scan, then a deep classifier (a DenseNet model in the 201 configuration \cite{huang2017densely}) processes the segmented lung areas to identify if the slice shows signs of COVID-19 virus.

\begin{figure*}[hbt!]
    \centering
    \includegraphics[width=\textwidth]{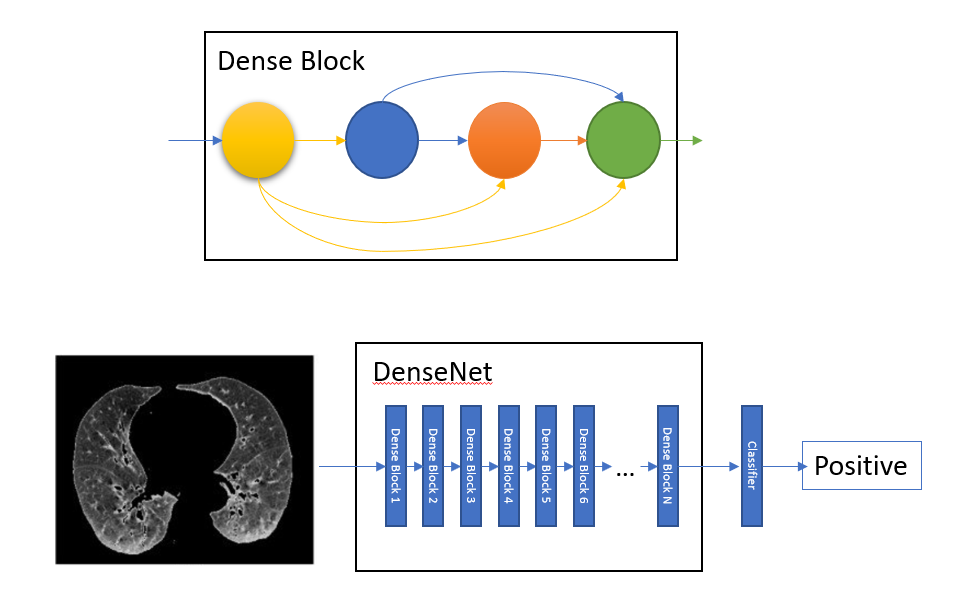}
        \centering
    \caption{The DenseNet architecture. Convolutional processing layers are grouped in Dense Blocks (\textit{top}). Features extracted in previous layers are concatenated and fed to all the next layers in the same Dense Block ensuring maximum information flow. Given that feature maps from previous layers are passed to the next layers, redundancy is avoided (i.e., later layers do not need to learn almost identical information from the immediately previous ones). In this way, each successive layer adds only a small number of feature maps, the so called \textit{growth factor}, thus requiring fewer parameters to achieve state-of-the-art performance. Multiple Dense Blocks can be concatenated and form a deeper network (\textit{bottom}).}
    \label{fig:dense}
\end{figure*}

Once the COVID-19 identification model is trained, we attempt to understand what features it employs to discriminate between positive and negative cases. Thus, to interpret the decisions made by the trained model we compute class-discriminative localization maps that attempt to provide visual explanations of the most significant input features for each class. To accomplish this we employ GradCAM~\cite{8237336} combined to VarGrad~\cite{NIPS2018_8160}.
More specifically, GradCAM is a technique to produce such interpretability maps through by investigating output gradient with respect to
feature map activations. More specifically, GradCAM generates  class-discriminative
localization map for any class $c$ by first computing the gradient of the score for class $c$, $s^c$, w.r.t feature activation maps $A_k$ of a given convolutional layer. Such gradients are then global-average-pooled to obtain the activation importance weights $w$, i.e.:
\begin{equation}
w^c_k = \sum_i \sum_j 	\frac{\partial y^c}{\partial A^k_{ij}}
\end{equation}

Afterwards, the saliency map $S^c$, that provides an overview of the activation importance for the class $c$, is computed through a weighted combination of activation maps, i.e.:

\begin{equation}
S^c = ReLU \left(\sum_k  w_k^cA^k \right)
\end{equation}

VarGrad is a technique used in combination to GradGAM and consists in performing multiple activation map estimates by adding, each time, Gaussian noise to the input data and then aggregating the estimates by computing the variance of the set.

\subsection{COVID-19 lesion identification and categorization}
\begin{figure*}
    \centering
    \includegraphics[width=\textwidth]{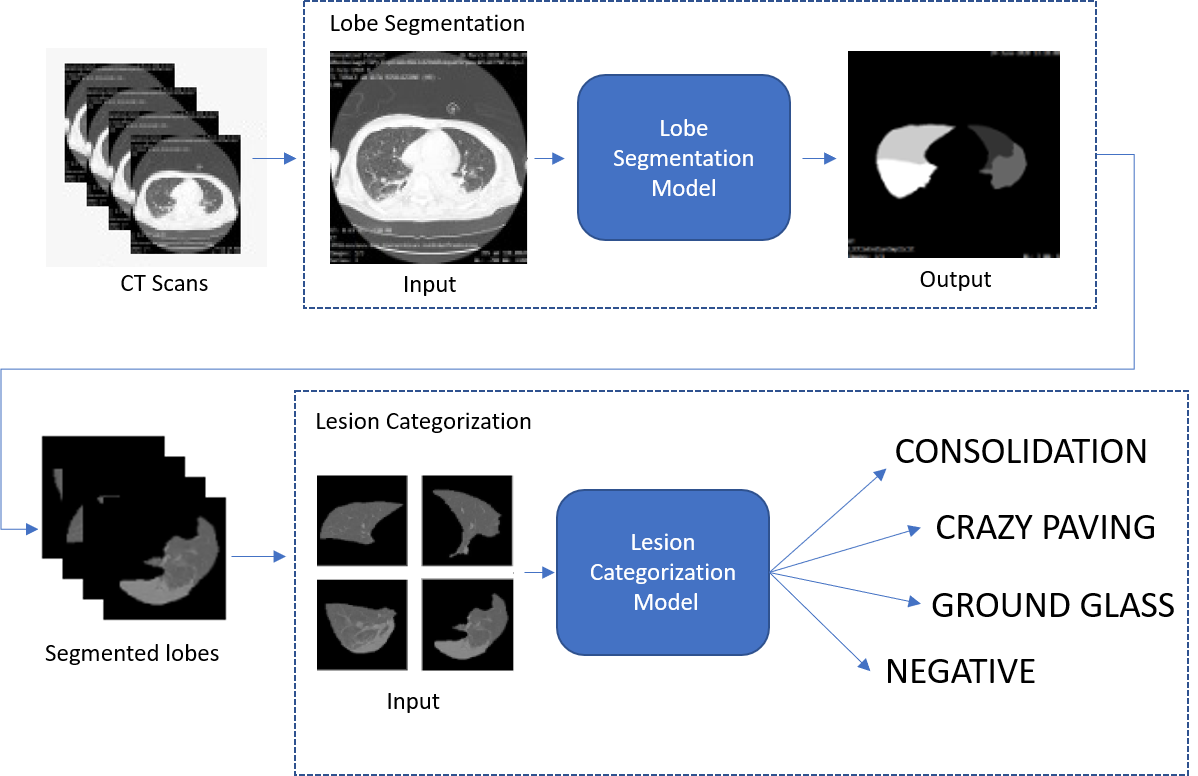}
    \caption{Overview of \textbf{COVID-19 lesion categorization} approach.}
    \label{fig:overview_lung}
\end{figure*}

An additional deep network activates only if the previous system identifies a COVID-19 positive CT scan. In that case, it works on the subset of slices identified as COVID-19 positives by the first AI system with the goal to localize and identify specific lesions (consolidation, crazy paving and ground glass). More specifically, the lesion identification system works on segmented lobes to seek COVID-19 specific patterns. The subsystem for lesion categorization employs the knowledge already learned by the COVID-19 detection module (shown in Fig. \ref{fig:overview}) and refines it for specific lesion categorization. An overview of the whole system is given in Fig. \ref{fig:overview_lung}.

\subsection{A Web-based Interface for Explaining AI decisions to Radiologists}
\begin{figure*}
    \centering
    \includegraphics[width=\textwidth]{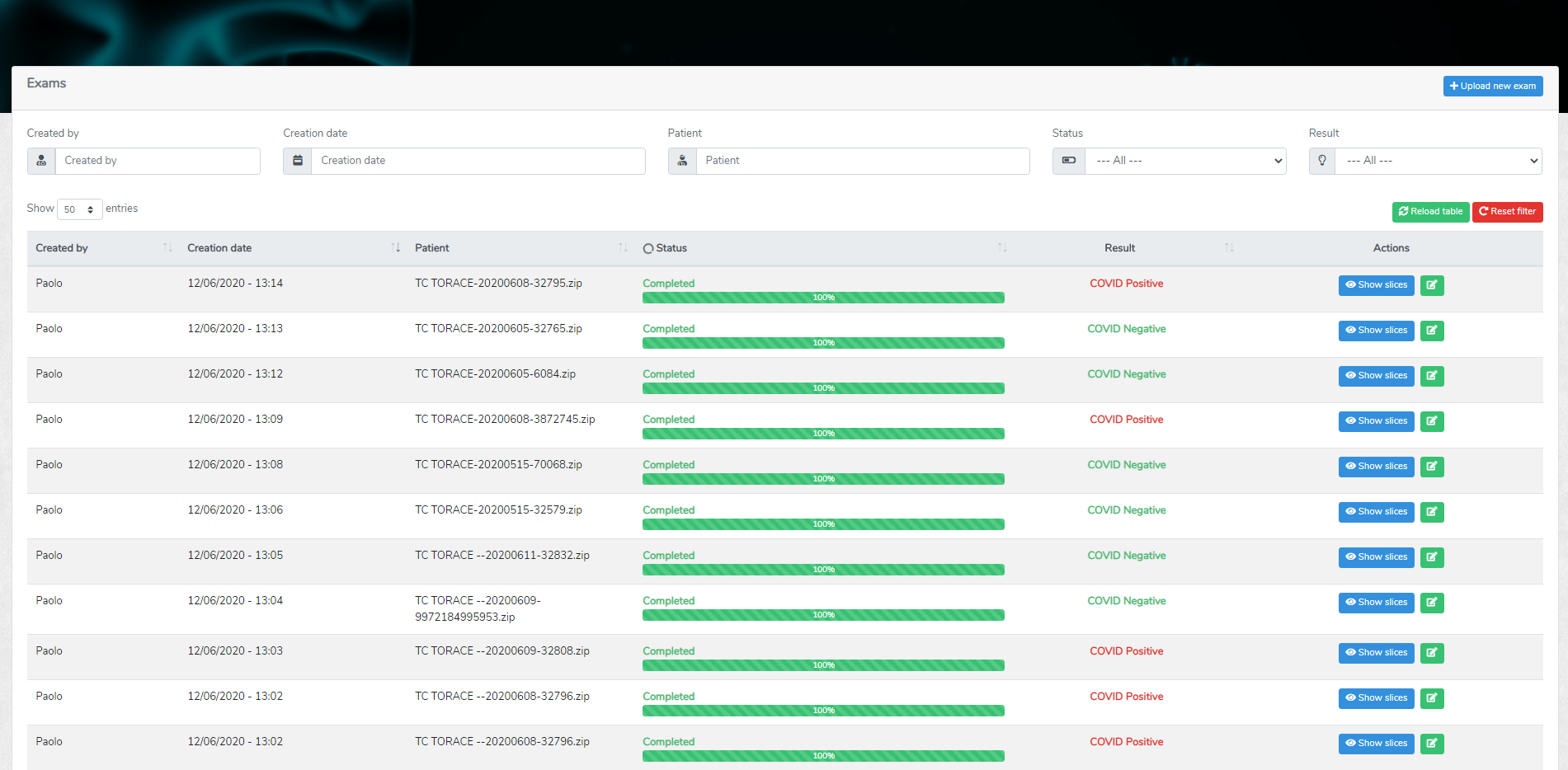}
    \caption{The main page of the AI-empowered web GUI for explainable AI. The user is presented with a list of the CT scan classifications reporting the models’ prediction.}
    \label{fig:main}
\end{figure*}

\begin{figure*}
    \centering
    \includegraphics[width=\textwidth]{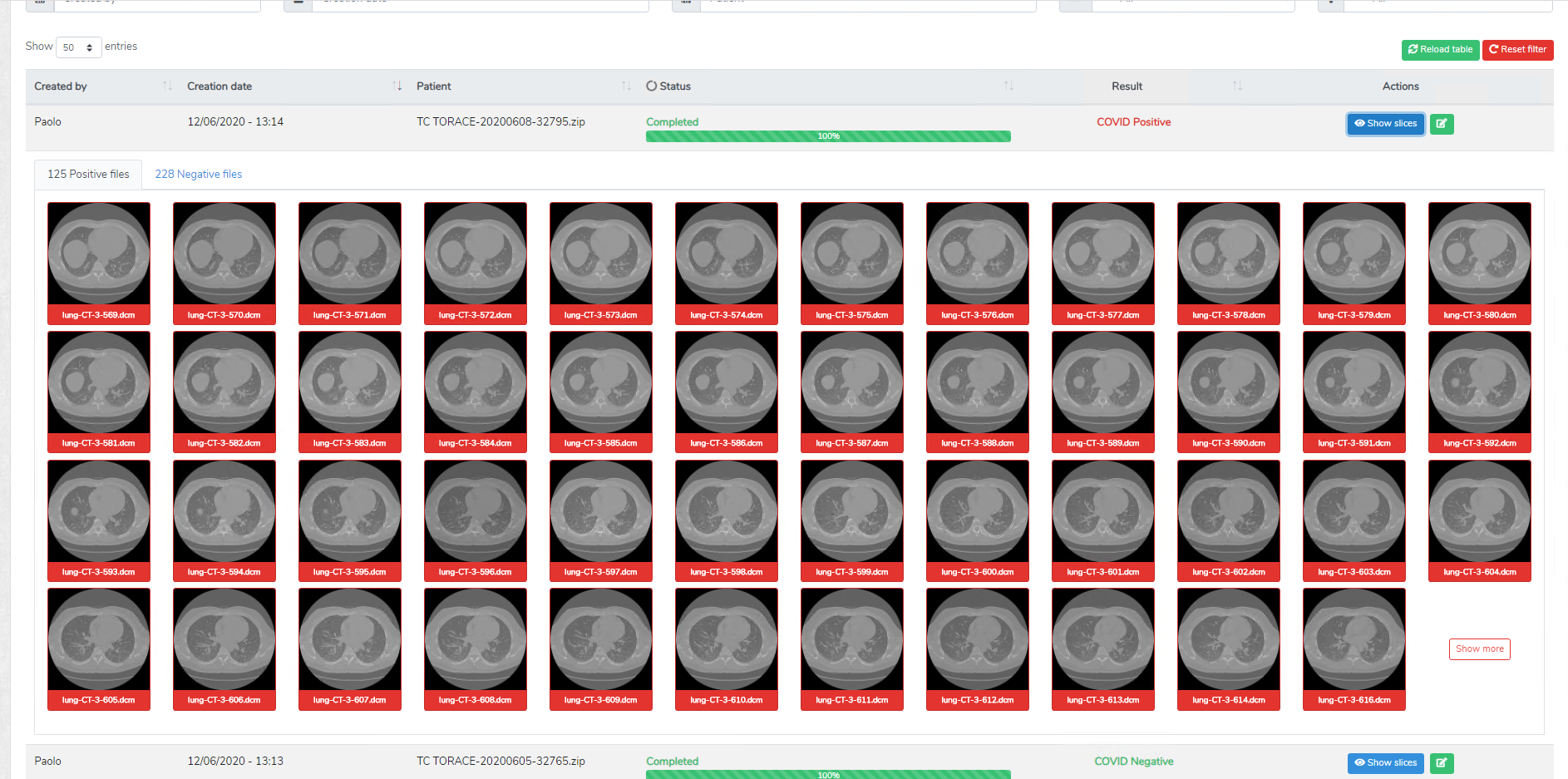}
    \caption{The summarized classification result showing the CT slices that the neural network classified as positive or negative.}
    \label{fig:summary}
\end{figure*}

In order to explain to radiologists, the decisions made by a “black-box” AI system, we integrated the inference pipeline for COVID-19 detection into a web-based application. The application was designed to streamline the whole inference process with just a few clicks and visualize the results with a variable grade of detail (Fig.~\ref{fig:main}). 
If the radiologists desire to see which CT slices were classified as positive or negative, they can click on “Show slices” where a detailed list of slices and their categorization is showed (Fig.~\ref{fig:summary}). 

\begin{figure*}
    \centering
    \includegraphics[width=\textwidth]{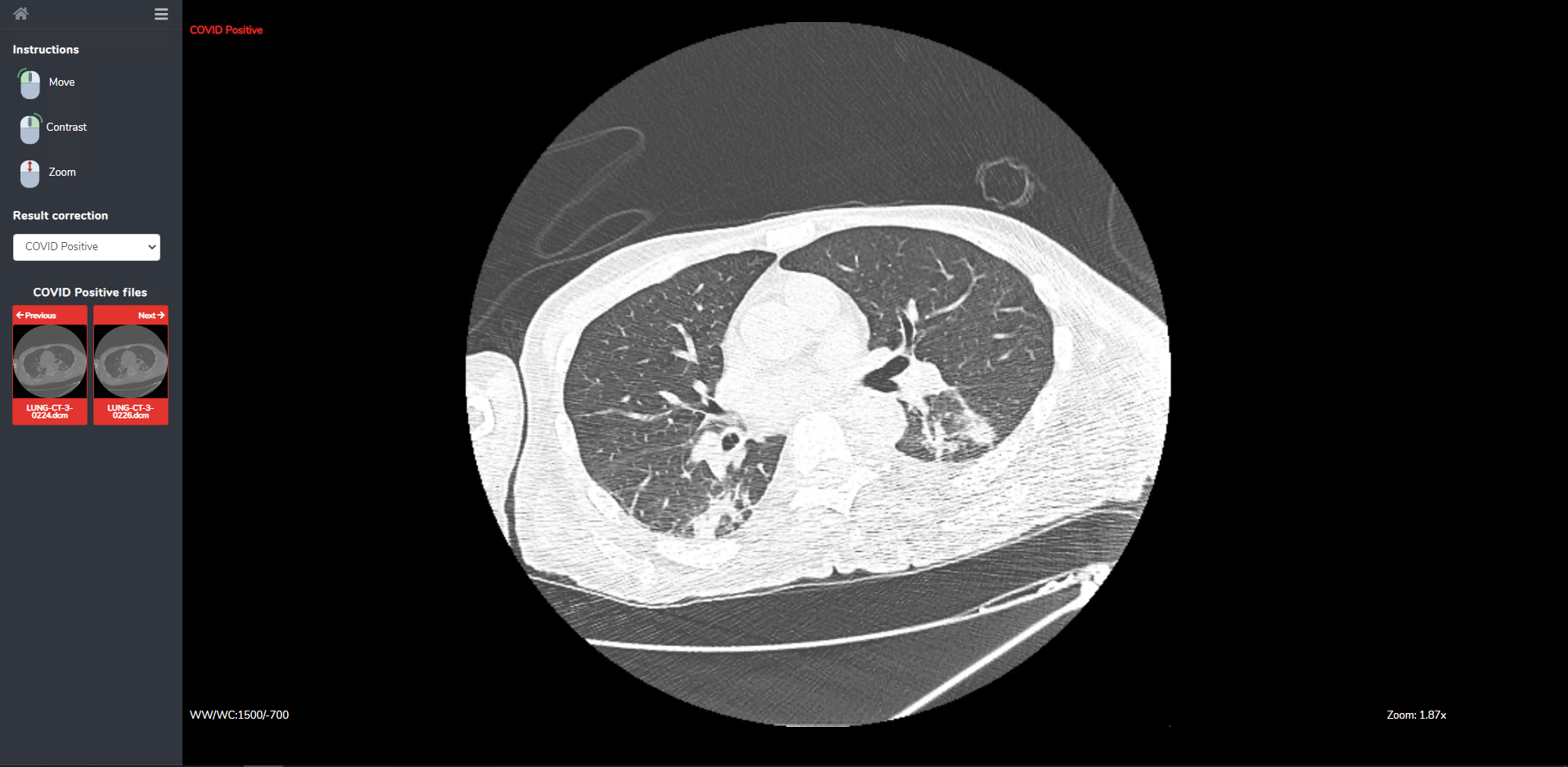}
    \caption{The slice inspection screen. In this screen the user can inspect each single slice and the AI models’ decisions.}
    \label{fig:inspect}
\end{figure*}

Because the models may not achieve perfect accuracy, a single slice inspection screen is provided, where radiologists can inspect more closely the result of the classification. It also features a restricted set of image manipulation tools (move, contrast, zoom) for aiding the user to make a correct diagnosis (Fig.~\ref{fig:inspect}).

The AI-empowered web system integrates also a relevance feedback mechanism where radiologists can correct the predicted outputs, and the AI module exploits such a feedback to improve its future assessments. Indeed, both at the CT scan level and at the CT slice level, radiologists can correct models’ prediction. The AI methods will then use the correct labels to enhance their future assessments.

%% file: segm.tex
Our lung-lobe segmentation model is based on the \emph{Tiramisu} network~\cite{jegou2017one}, a fully-convolutional DenseNet~\cite{huang2017densely} in a U-Net architecture~\cite{ronneberger2015u}. The model consists in two data paths: the downsampling one, that aims at extracting features and the upsampling one that aims at generating the output images (masks). Skip connections (i.e., connections starting from a preceding layer in the network's pipeline to another one found later bypassing intermediate layers) aim at propagating high-resolution details by sharing feature maps between the two paths. 

In this work,  our segmentation model follows the Tiramisu architecture, but with two main differences:

\begin{itemize}
    \item Instead of processing each single scan individually, convolutional LSTMs~\cite{xingjian2015convolutional} are employed at the network's bottleneck layer to exploit the spatial axial correlation of consecutive scan slices.
    \item In the downsampling and upsampling paths, we add residual squeeze-and-excitation layers~\cite{hu2017squeeze}, in order to emphasize relevant features and improve the representational power of the model.
\end{itemize}
    
\begin{figure*}
	\centering
	\includegraphics[width=1\textwidth]{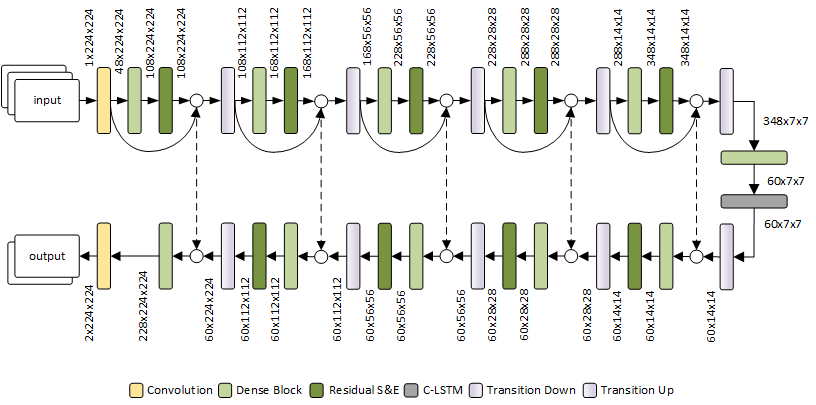}
	\caption{The proposed segmentation architecture, consisting of a downsampling path (\textit{top}) and an upsampling path (\textit{bottom}), interconnected by skip connections and by the bottleneck layer.}
	\label{fig:segmentation_arch}
\end{figure*}

Before discussing the properties and advantages of the above modifications, we first introduce the overall architecture, shown in Fig.~\ref{fig:segmentation_arch}.

The input to the model is a sequence of 3 consecutive slices -- suitably resized to 224$\times$224 -- of a CT scan, which are processed individually and combined through a convolutional LSTM layer. Each slice is initially processed with a standard convolutional layer to expand the feature dimensions. The resulting feature maps then go through the downsampling path of the model (the encoder) consisting of five sequences of dense blocks, residual squeeze-and-excitation layers and transition-down layers based on max-pooling. In the encoder, the feature maps at the output of each residual squeeze-and-excitation layer are concatenated with the input features of the preceding dense block, in order to encourage feature reuse and improve their generalizability. At the end of the downsampling path, the \emph{bottleneck} of the model consists of a dense block followed by a convolutional LSTM. The following upsampling path is symmetric to the downsampling one, but it features: 1) skip connections from the downsampling path for concatenating feature maps at the corresponding layers of the upsampling path; 2) transition-up layers implemented through transposed convolutions. Finally, a convolutional layer provides a 6-channel segmentation map, representing, respectively, the log-likelihoods of the lobes (5 channels, one for each lobe) and non-lung (1 channel) pixels. 

In the following, we review the novel characteristics of the proposed architecture.\\

\noindent \textbf{Residual squeeze-and-excitation layers.} Explicitly modeling interdependencies between feature channels has demonstrated to enhance performance of deep architectures; squeeze-and-excitation layers~\cite{hu2017squeeze} instead aim to select informative features and to suppress the less useful ones. In particular, a set of input features of size $C\times H \times W$ is squeezed through average-pooling to a $C \times 1 \times 1$ vector, representing global feature statistics. The ``excitation'' operator is a fully-connected non-linear layer that translates the squeezed vector into channel-specific weights that are applied to the corresponding input feature maps.\\

\noindent \textbf{Convolutional LSTM.} We adopt a recurrent architecture to process the output of the bottleneck layer, in order to exploit the spatial axial correlation between subsequent slices and enhance the final segmentation by integrating 3D information in the model. Convolutional LSTMs~\cite{xingjian2015convolutional} are commonly used to capture spatio-temporal correlations in visual data (for example, in videos), by extending traditional LSTMs using convolutions in both the \textit{input-to-state} and the \textit{state-to-state} transitions.
Employing recurrent convolutional layers allows the model to take into account the context of the currently-processed slice, while keeping the sequentiality and without the need to process the entire set of slices in a single step through channel-wise concatenation, which increases feature sizes and loses information on axial distance. 

\begin{figure*}
    \centering
    \includegraphics[width=1\textwidth]{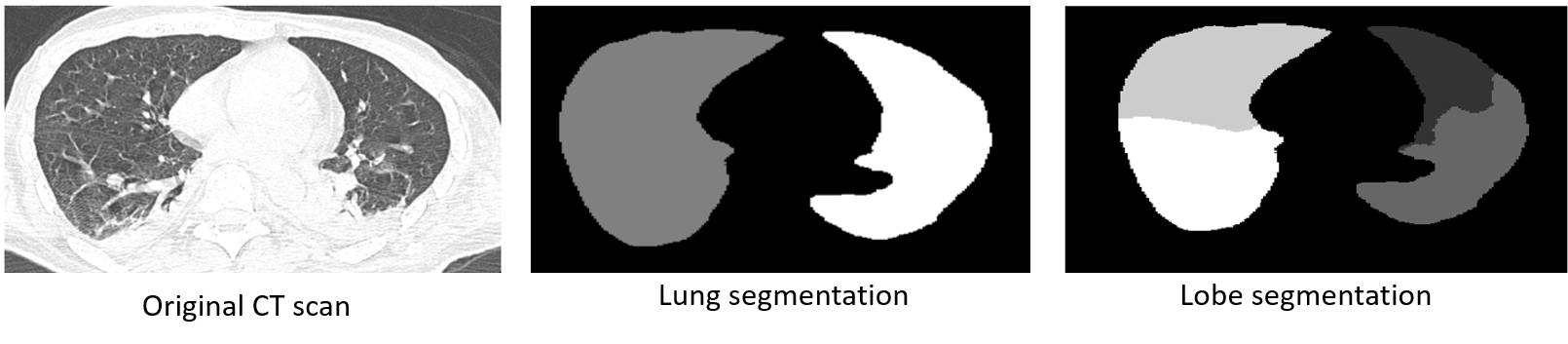}
    \caption{Example of lung and lobes segmentation.}
    \label{fig:segmentation}
\end{figure*}

Fig. \ref{fig:segmentation} shows an example of automated lung and lobe segmentation from a CT scan by employing the proposed segmentation network.
The proposed segmentation network is first executed on the whole CT scan for segmenting only lung (and lobes); the segmented CT scan is then passed to the downstream classification modules for COVID-19 identification and lesion categorization.

%% file: results.tex
\subsection{Dataset}

Our dataset contains 72 CT scans of COVID-19 positive patients (positivity confirmed both by a molecular - reverse transcriptase–polymerase chain reaction for SARS-coronavirus RNA from nasopharyngeal aspirates - and an IgG or IgM antibody test) and 94 CT scans of COVID-19 negative subjects (35 patients with interstitial pneumonia but tested negative to COVID-19 and  59 controls). 

CT scans were performed on a multi-detector row helical CT system scanner \footnote{Bright Speed, General Electric Medical Systems, Milwaukee, WI} using 120 kV pp, 250 mA, pitch of 1.375, gantry rotation time of 0,6 s and time of scan 5,7 s. The non-contrast scans were reconstructed with slice thicknesses of 0.625 mm and spacing of 0.625 mm with high-resolution lung algorithm. The images obtained on lung (window width, 1,000–1,500 H; level, –700 H) and mediastinal (window width, 350 H; level, 35–40 H) settings were reviewed on a picture archiving and communication system workstation\footnote{Impax ver. 6.6.0.145, AGFA Gevaert SpA, Mortsel, Belgium}. 
CT scans of positive patients were also annotated by three expert radiologists (through consensus) who selected a subset of slices and annotated them with the type  (Consolidation, Ground Glass and Crazy Paving) and the location (combinations of left/right/central and posterior/anterior) of the lesion. In total about 2,400 slices were annotated with COVID-19 lesions and about 3,000 slices of negative patients with no lesion. 
Tab.~\ref{tab:data} provides an overview of all the CT scans and annotations in our dataset. 

For training the lung/lobe segmentation model we adopted a combination of the LIDC~\cite{a4d01420decb47e5ac928fa63a199de0},  LTRC\footnote{https://ltrcpublic.com/} and ~\cite{hofmanninger2020automatic} datasets, for a total of 300 CT scans. Annotations on lung/lobe areas were done manually by three expert radiologists.

\begin{table}
\centering

    \begin{tabular}{ c c c }
    \toprule
    \multicolumn{3}{c}{Dataset statistics}\\
    \midrule
    \midrule
    \multicolumn{3}{c}{\textbf{CT Data}}\\
    \midrule
    \midrule
        CT Scans &  & 166\\
    \midrule
    & COVID-19+ & 72\\
    & COVID-19- & 94\\
    \midrule
    \midrule
    \multicolumn{3}{c}{\textbf{Annotations}}\\
    \midrule
    Positive slices	& & 2,390\\
    \midrule
     & Ground Glass		& 1,035\\
    & Crazy Paving	& 	757\\
    & Consolidation	& 	598\\
    \midrule
    Negative slices	&  &2,988\\
    \midrule
    \midrule

    \end{tabular}
    \caption{CT Dataset for training and testing of the AI models.}
    \label{tab:data}
\end{table}

\subsection{Training Procedure}

The COVID-19 detection network is a DenseNet201, which was used pretrained on the ImageNet dataset~\cite{deng2009imagenet}. The original classification layers in DenseNet201 were replaced by a 2-output linear layer for the COVID-19 positive/negative classification. 
Among the set of 166 CT scans, we used 95 scans (36 positives and 59 negatives) for training, 9 scans for validation (5 positives and 4 negatives) and 62 scans (31 positives and 31 negatives) for test. To compare the AI performance to the human one, the test set of 62 CT scans was provided to three expert radiologists for blind evaluation.
Given the class imbalance in the training set, we used the weighted binary cross-entropy (defined in \ref{eq:bloss}) as training loss and RT-PCR virology test as training/test labels. 

The weighted binary cross-entropy loss for a sample classified as $x$ with target label $y$ is then calculated as:
\begin{equation}
WBCE = - w \left[ y \cdot \log x + (1 - y) \cdot \log (1 - x) \right]
\label{eq:bloss}
\end{equation}

where $w$ is defined as the ratio of the number negative samples to the total number of samples if the label is positive and vice versa. This way the loss results higher when misclassifying a sample that belongs to the less frequent class.
It is important to highlight that splitting refers to the entire CT scan and not to the single slices: we made sure that full CT scans were not assigned in different splits to avoid any bias in the performance analysis. This is to avoid the deep models overfit the data by learning spurious information from each CT scan, thus invalidating the training procedure, thus enforcing robustness to the whole approach.
Moreover, for the COVID-19 detection task, we operate at the CT level by processing and categorizing each single slice. To make a decision for the whole scan, we perform voting: if 10\% of total slices is marked as positive then the whole exam is considered as a COVID-19 positive, otherwise as COVID-19 negative. The choice of the voting threshold was done empirically to maximize training performance.

The lesion categorization deep network is also a DenseNet201 model where classification layers were replaced by a 4-output linear layer (\textit{ground glass, consolidation, crazy paving, negative}). 
The lesion categorization model processes lobe segments (extracted by our segmentation model) with the goal to identify specific lesions. Our dataset contains 2,488 annotated slices; in each slice multiple lesion annotations with relative location (in lobes) are available. Thus, after segmenting lobes from these images we obtained 5,264 lobe images. We did the same on CT slices of negative patients (among the 2,950 available as shown in Tab.~\ref{tab:data}) and selected 5,264 lobe images without lesions. Thus, in total, the the entire set consisted of 10,528 images. We also discarded the images for which lobe segmentation produced small regions indicating a failure in the segmentation process. 
We used a fixed test split consisting of 195 images with consolidation, 354 with crazy paving, 314 with ground glass and 800 images with no lesion. The remaining images were split into training and validation sets with the ratio 80/20. Given the class imbalance in the training set, we employed  weighted cross-entropy as training loss. 

The weighted cross-entropy loss for a sample classified as $x$ with target label $y$ is calculated as:

\begin{equation}
WCE = -w\sum^Cy \cdot log(x)
\label{eq:celoss}
\end{equation}

where $C$ is the set of all classes. The weight $w$ for each class $c$ is defined as:

\begin{equation}
w_c = \frac{N - Nc}{N}
\label{eq:weight}
\end{equation}

where $N$ is the total number of samples and $N_c$ is the number of samples that have label $c$.

Since the model is the same as the COVID identification network, i.e., DenseNet201, we started from the network trained on the COVID-identification task and fine-tune it on the categorization task to limit overfitting given the small scale of our dataset. 

For both the detection network and the lesion categorization network, we used the following hyperparameters: batch-size = 12, learning rate = 1e-04, ADAM back-propagation optimizer with beta values 0.9 and 0.999, eps = 1e-08 and weight decay = 0 and the back-propagation method was used to update the models’ parameters during training. 
Detection and categorization networks were trained for 20 epochs.
In both cases, performance are reported at the highest validation accuracy.

For lung/lobe segmentation, input images were normalized to zero mean and unitary standard deviation, with statistics computed on the employed dataset. In all the experiments for our segmentation model, input size was set to $224 \times 224$, initial learning rate to 0.0001, weight decay to 0.0001 and batch size to 2, with RMSProp as optimizer. When C-LSTMs were employed, recurrent states were initialized to zero and the size of the input sequences to the C-LSTM layers was set to 3. Each training was carried out for 50 epochs.

\subsection{Performance Evaluation}
In this section report the performance of the proposed model for lung/lobe segmentation, COVID-19 identification and lesion categorization.

\subsubsection{Lobe segmentation}

Our segmentation model is based on the Tiramisu model~\cite{jegou2017one} with the introduction of \emph{squeeze-and-excitation} blocks and of a convolutional LSTM (either unidirectional or bidirectional) after the bottleneck layer. 
In order to understand the contribution of each module, we first performed ablation studies by testing the segmentation performance of our model using different architecture configurations:
\begin{itemize}
 \item Baseline: the vanilla Tiramisu model described in~\cite{jegou2017one};
 \item Res-SE: residual \emph{squeeze-and-Excitation} module are integrated in each dense block of the Tiramisu architecture;
 \item C-LSTM: a unidirectional convolutional LSTM is added after the bottleneck layer of the Tiramisu architecture;
 \item Res-SE + C-LSTM: variant of the Tiramisu architecture that includes both residual \emph{squeeze-and-Excitation} at each dense layer and a unidirectional convolutional LSTM after the bottleneck layer.
\end{itemize}

We also compared the performance against the U-Net architecture proposed in \cite{hofmanninger2020automatic} that is largely adopted for lung/lobe segmentation.

All architectures were trained for 50 epochs by splitting the employed lung datasets into a training, validation and test splits using the 70/10/20 rule. Results in terms of Dice score coefficient (DSC) are given in Tab.~\ref{tab:model_selection}.
It has to noted that unlike ~\cite{hofmanninger2020automatic}, we computed DSC on all frames, not only on the lung slices.\\
The highest performance is obtained with the Res-SE + C-LSTM configuration, i.e., when adding \emph{squeeze-and-excitation} and the unidirectional C-LSTM at the bottleneck layer of the Tiramisu architecture. This results in an accuracy improvement of over 4 percent points over the baseline. In particular, adding \emph{squeeze-and-excitation} leads to a 2 percent point improvement over the baseline. 
Segmentation results are computed using data augmentation obtained by applying random affine transformations (rotation, translation, scaling and shearing) to input images. 
The segmentation network is then applied to our COVID-19 dataset for prior segmentation without any additional fine-tuning to demonstrate also its generalization capabilities.

\begin{table}
	\centering
	{
		\begin{tabular}{ccc}
			\toprule
			\textbf{Model} & \textbf{Lung segmentation} & \textbf{Lobe segmentation}  \\
			\midrule
			Baseline Tiramisu \cite{jegou2017one} & $89.41 \pm 0.45$ & 77.97 $\pm$ 0.31\\ 
			Baseline + Res-SE  			& $91.78 \pm 0.52$ & 80.12 $\pm$ 0.28 \\
			Baseline + C-LSTM 				& $91.49 \pm 0.57$ & 79.47 $\pm$ 0.38\\
			Baseline + Res-SE + C-LSTM 	& \textbf{94.01 $\pm$ 0.52} &\textbf{ 83.05 $\pm$ 0.27}\\
			\bottomrule
		\end{tabular}
			\vspace{5pt}
		\caption{Ablation studies of our segmentation network in terms of dice score. Best results are shown in bold. Note: we did not compute confidence intervals on these scores as they are obtained from a very large set of CT pixels.}
		\label{tab:model_selection}
	}
\end{table}

\subsubsection{COVID-19 assessment}

We compute results both for COVID-19 detection and lesion categorization and compare to those yielded by three experts with different degree of expertise:
\begin{enumerate}
    \item Radiologist 1: a physician expert in thoracic radiology ($\sim$30 years of experience) with over 30,000 examined CT scans;
    \item Radiologist 2: a physician expert in thoracic radiology ($\sim$10 years of experience) with over 9,000 examined CT scans;
    \item Radiologist 3: a resident student in thoracic radiologist ($\sim$3 years of experience) with about 2,000 examined CT scans.
\end{enumerate}
We also assess the role of prior segmentation on the performance. This means that in the pipelines showed in Figures \ref{fig:overview} and \ref{fig:overview_lung} we removed the segmentation modules and performed classification using the whole CT slices using also information outside the lung areas. Results for COVID-19 detection are measured in terms of sensitivity and specificity and given in Tables \ref{tab:sens} and \ref{tab:spec}.

\begin{table}
\centering
    \begin{tabular}{ c c c}
    \toprule
     & Sensitivity & C.I. (95\%)\\
    \midrule
    Radiologist 1	& 83.9\% & [71.8\% -- 91.9\%]\\
    Radiologist 2   & 87.1\% & [75.6\% -- 94.3\%]\\
    Radiologist 3   & 80.6\% & [68.2\% -- 89.5\%]\\
    \midrule
    AI Model without lung segmentation &	83.9\% & [71.8\% -- 91.9\%]\\
    AI Model with lung segmentation	& \textbf{90.3}\% &\textbf{ [79.5\% -- 96.5\%]}\\
    \bottomrule
 \hline
    
    \end{tabular}
    \caption{Sensitivity (together with 95\% confidence interval) comparison between manual readings of expert radiologists and the AI model for COVID-19 detection without lung segmentation and AI model with segmentation.}
    \label{tab:sens}
\end{table}

Thus, the AI model using lung segmentation achieves the best performance outperforming expert radiologists in the COVID-19 assessment. Furthermore, performing lung segmentation improves by about 6 percent points both the sensitivity and the specificity, demonstrating its effectiveness. The important aspect to highlight is that expert radiologists during the annotation process did not have to segment lungs or lobes, showing the generalization capabilities of the proposed deep learning-based methods.

\begin{table}
\centering
    \begin{tabular}{ccc}
    \toprule
     & Specificity & C.I. (95\%)\\
    \midrule
    Radiologist 1	& 87.1\% & [75.6\% -- 94.3\%]\\
    Radiologist 2 &	87.1\% & [75.6\% -- 94.3\%]\\
    Radiologist 3 &	90.3\% &  [79.5\% -- 96.5\%]\\
    \midrule
    AI Model without lung segmentation &	87.1\%  & [75.6\% -- 94.3\%]\\
    AI Model with lung segmentation	& \textbf{93.5}\% &\textbf{[83.5\% -- 98.5\%]}\\
    \bottomrule
     \hline
    \end{tabular}
    \caption{Specificity (together with 95\% confidence interval) comparison between manual readings of expert radiologists and the AI model for COVID-19 detection without lung segmentation and AI model with segmentation.}
    \label{tab:spec}
\end{table}

As a backbone model for COVID-19 identification, we employed DenseNet201 since it yielded the best performance when compared to other state of the art models, as shown in Table~\ref{tab:comparison}.
In all the tested cases, we used upstream segmentation through the model described in Sect.~\ref{ssec:segmentation}. Voting threshold was set to 10\% on all cases.

\begin{table}
\centering
    \begin{tabular}{ccccc}
    \toprule
    Model & Variant & Sensitivity (CI) & Specificity (CI)  & Accuracy (CI)\\
    \midrule
    AlexNet & -- &71.0\% \footnotesize{(57.9--81.6)} &90.3\% \footnotesize{(79.5--96.5)} & 80.7\% \footnotesize{(68.3--89.5)} \\
    \midrule
    \multirow{4}{*}{ResNet} &18  & 71.0\% \footnotesize{(57.9--81.6)}&  93.5\% \footnotesize{(83.5--98.5)}& 82.3\%
    \footnotesize{(70.1--90.7)}\\
     &34  & 80.7\% \footnotesize{(68.3--89.5)} & 90.3\% \footnotesize{(79.5--96.5)}  & 85.5\% \footnotesize{(73.7--93.1)}\\
     &50  & 83.9\% \footnotesize{(71.9--91.9)} & 90.3\% \footnotesize{(79.5--96.5)} & 87.1\% \footnotesize{(75.6--94.3)}\\
     &101  &77.4\% \footnotesize{(64.7--89.9)} & 87.1\% \footnotesize{(75.6--94.3)} & 82.3\% \footnotesize{(70.1--90.7)}\\
     &152  & 77.4\% \footnotesize{(64.7--89.9)}&  90.3\% \footnotesize{(79.5--96.5)} &83.9\% \footnotesize{(71.9--91.9)}\\
    \midrule
    \multirow{2}{*}{DenseNet} & 121 &77.4\% \footnotesize{(64.7--89.9)} &93.5\% \footnotesize{(83.5--98.5)}  &85.5\%  \footnotesize{(73.7--93.1)}\\
     & 169 &67.9\% \footnotesize{(83.5--98.5)} &93.5\% \footnotesize{(83.5--98.5)}  &81.4\% \footnotesize{(68.7--90.2)}\\
     & 201 &\textbf{90.3\% \footnotesize{(79.5--96.5)}} &\textbf{93.5\% \footnotesize{(83.5--98.5)}}  &\textbf{91.9\% \footnotesize{(81.5--97.5)}}\\
    \midrule
    SqueezeNet & -- &66.7\% \footnotesize{(54.5--78.9)}  &93.5\% \footnotesize{(83.5--98.5)}  &81.4\% \footnotesize{(68.7--90.2)} \\
    \midrule
    ResNeXt & -- &77.4\% \footnotesize{(64.7--86.9)} &90.3\% \footnotesize{(79.5--96.5)}  &83.9\% \footnotesize{(71.9--91.9)}\\
    \bottomrule
     \hline
    \end{tabular}
    \caption{COVID-19 classification accuracy by several state of the art models. Values in parentheses indicate 95\% confidence intervals (CI).}
    \label{tab:comparison}
\end{table}

\begin{figure*}
    \centering
    \includegraphics[width=\textwidth]{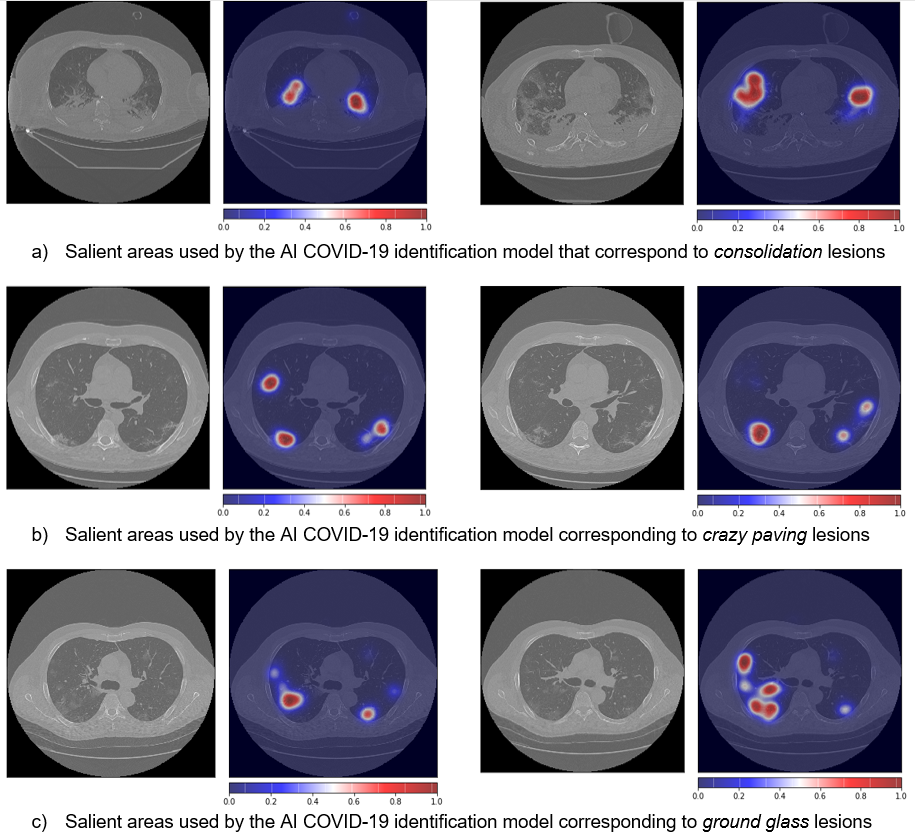}
    \caption{Lung salient areas identified automatically by the AI model for CT COVID-19 identification.}
    \label{fig:salient}
\end{figure*}

In order to enhance trust in the devised AI models, we analyzed what features these methods employ for making the COVID-19 diagnosis decision. This is done by investigating which artificial neurons fire the most, and then projecting this information to the input images. To accomplish this we combined GradCAM~\cite{8237336} with VarGrad~\cite{NIPS2018_8160}\footnote{https://captum.ai/} and, Fig.~\ref{fig:salient} shows some examples of the saliency maps generated by interpreting the proposed AI COVID-19 classification network. It is interesting to note that the most significant activation areas correspond to the three most common lesion types, i.e., ground glass, consolidation and crazy paving. This is remarkable as the model has indeed learned the COVID-19 peculiar patterns without any information on the type of lesions (to this end, we recall that for COVID-19 identification we only provide, at training times, the labels “positive” or “negative”, while no information on the type of lesions is given). 

For COVID-19 lesion categorization we used mean (and per-class) classification accuracy over all lesion types and per lesion that are provided, respectively, in Table \ref{tab:acc}.

\begin{table*}[h!]
\centering
    \begin{tabular}{ c c c}
    \toprule
    	& 	Model no\_segm	& 	Model w\_segm\\
    \midrule
    Consolidation & 77.8\% \footnotesize{(69.9--84.1)} & 97.9\% \footnotesize{(93.6--99.8)}\\
    Ground glass  & 18.6\% \footnotesize{(14.1--24.1)} & 41.3\% \footnotesize{(35.1--47.7)}\\
    Crazy Paving & 57.1\% \footnotesize{(49.4--64.4)}& 98.3\%  \footnotesize{(94.8--99.8)}\\
    Negative   & 99.3\% \footnotesize{(98.6--99.7)} & 99.9\% \footnotesize{(99.5--100)}\\
    \midrule
    Average & 63.2\%& 84.4\%\\
    \bottomrule
 \hline
    
    \end{tabular}
    
    \caption{Per-class accuracy for lesion categorization between manual readings of expert radiologists and the AI model without lung segmentation and AI model with segmentation. Values in parentheses indicate 95\% confidence intervals (CI).}
    \label{tab:acc}
\end{table*}

Mean lesion categorization accuracy reaches, when operating at the lobe level, about 84\% of performance. The lowest performance is obtained on ground glass, because ground glass opacities are specific CT findings that can appear also in normal patients with respiratory artifact. 
Operating at the level of single lobes yields a performance enhancement of over 21 percent points, and, also in this case, radiologists did not have to perform any lobe segmentation annotation, reducing significantly their efforts to build AI models. The most significant improvement when using lobe segmentation w.r.t. no segmentation is obtained \textit{Crazy Paving} class, i.e., 98.3\% against 57.1\%.

Despite the CT diagnosis of COVID-19 pneumonia seems an easy task for experienced radiologists, the results show that our system is able to outperform them providing more accurate decisions. Artificial intelligence (AI), in particular, is able to identify more accurately lung lesions, in particular the smaller and undefined ones (as those highlighted in Fig. \ref{fig:salient}) The identification elements increases the sensitivity and specificity of the method for the correct diagnosis.
The results obtained both for COVID-19 identification and lesion categorization pave the way to further improvement by implementing an advanced COVID-19 CT/RX image-driven diagnostic pipeline interpretable and strongly robust to provide not only the diseases identification and differential diagnosis but also the risk of disease progression.

%% file: conclusions.tex
In this work we have presented an AI-based pipeline for automated lung segmentation, COVID-19 detection and COVID-19 lesion categorization from CT scans. Results showed a sensitivity of 90\% and a specificity of 93.5\% for COVID-19 detection and average lesion categorization accuracy of about 64\%. Results also show that a significant role is played by prior lung and lobe segmentation that allowed us to enhance performance of about 6 percent points. 
The AI models are then integrated into an user-friendly GUI to support AI explainability for radiologists, which is publicly available at \url{http://perceivelab.com/covid-ai}. To the best of our knowledge, this is the first AI-based software, publicly available, that attempts to explain radiologists what information is used by AI methods for making decision and that involve proactively in the loop to further improve the COVID-19 understanding.
These results pave the way to further improvement to provide not only the diseases identification and differential diagnosis but also the risk of disease progression.

%% file: ack.tex
We thank the “Covid 19 study group” from Spallanzani Hospital (Maria Alessandra Abbonizio, Chiara Agrati, Fabrizio Albarello, Gioia Amadei, Alessandra Amendola, Mario Antonini, Raffaella Barbaro, Barbara Bartolini, Martina Benigni, Nazario Bevilacqua, Licia Bordi, Veronica Bordoni, Marta Branca, Paolo Campioni, Maria Rosaria Capobianchi, Cinzia Caporale, Ilaria Caravella, Fabrizio Carletti, Concetta Castilletti, Roberta Chiappini, Carmine Ciaralli, Francesca Colavita, Angela Corpolongo, Massimo Cristofaro, Salvatore Curiale, Alessandra D’Abramo, Cristina Dantimi, Alessia De Angelis, Giada De Angelis, Rachele Di Lorenzo, Federica Di Stefano, Federica Ferraro, Lorena Fiorentini, Andrea Frustaci, Paola Gallì, Gabriele Garotto, Maria Letizia Giancola, Filippo Giansante, Emanuela Giombini, Maria Cristina Greci, Giuseppe Ippolito, Eleonora Lalle, Simone Lanini, Daniele Lapa, Luciana Lepore, Andrea Lucia, Franco Lufrani, Manuela Macchione, Alessandra Marani, Luisa Marchioni, Andrea Mariano, Maria Cristina Marini, Micaela Maritti, Giulia Matusali, Silvia Meschi, Francesco Messina Chiara Montaldo, Silvia Murachelli, Emanuele Nicastri, Roberto Noto, Claudia Palazzolo, Emanuele Pallini, Virgilio Passeri, Federico Pelliccioni, Antonella Petrecchia, Ada Petrone, Nicola Petrosillo, Elisa Pianura, Maria Pisciotta, Silvia Pittalis, Costanza Proietti, Vincenzo Puro, Gabriele Rinonapoli, Martina Rueca, Alessandra Sacchi, Francesco Sanasi, Carmen Santagata, Silvana Scarcia, Vincenzo Schininà, Paola Scognamiglio, Laura Scorzolini, Giulia Stazi, Francesco Vaia, Francesco Vairo, Maria Beatrice Valli) for the technical discussion and critical reading of this manuscript.

%% file: consent.tex
All data and methods were carried out in accordance to the General Data Protection Regulation 2016/679. The experimental protocols were approved by the Ethics Committee of the National Institute for Infectious Diseases Lazzaro Spallanzani in Rome. All patients enrolled in the study were over 18 at the time of their participation in the experiment and signed informed consent.